\documentclass{article}
\usepackage{spconf,amsmath,graphicx,hyperref}
\usepackage{kotex}
\usepackage{xspace}
\usepackage{cleveref}
\usepackage{enumitem}
\usepackage{nicematrix}

\title{Hierarchical Dual-Graph Retrieval for Long-form Audio Reasoning}
\name{Eunsang Lee$^*$, Taein Ju\sthanks{Equal contribution. \; $^\dagger$Corresponding author.
}, Jiyoung Lee$^\dagger$}
\address{Division of Artificial Intelligence \& Software, Ewha Womans University}

\usepackage{tabularx}
\usepackage{amsmath}
\usepackage{makecell}
\usepackage{array} 
\usepackage{multirow}
\usepackage{lipsum}
\usepackage{booktabs}
\usepackage{color}
\usepackage{colortbl}
\usepackage{subcaption}
\usepackage{spverbatim}
\usepackage{listings}
\usepackage{cleveref}

\crefname{figure}{Fig.}{Figs.}
\Crefname{figure}{Fig.}{Figs.}

\crefname{table}{Tab.}{Tabs.}
\Crefname{table}{Tab.}{Tabs.}

\crefname{section}{Sec.}{Secs.}
\Crefname{section}{Sec.}{Secs.}

\crefname{equation}{Eq.}{Eqs.}
\Crefname{equation}{Eq.}{Eqs.}

\crefname{algorithm}{Alg.}{Algs.}
\Crefname{algorithm}{Alg.}{Algs.}

\definecolor{codekeyword}{rgb}{0.0, 0.0, 0.5}   % Dark blue for keywords
\definecolor{codecomment}{rgb}{0.0, 0.5, 0.0}   % Dark green for comments
\definecolor{codestring}{rgb}{0.56, 0.0, 1.0}   % Purple for strings
\definecolor{backcolour}{rgb}{0.98,0.98,0.97}

\definecolor{lastcol}{gray}{0.95} 

\lstdefinestyle{pythonstyle}{
    language=Python,                          % Set language to Python
    basicstyle=\ttfamily\small,               % Basic font style
    keywordstyle=\color{codekeyword}\bfseries,% Keywords in bold dark blue
    commentstyle=\color{codecomment}\itshape, % Comments in italic dark green
    stringstyle=\color{codestring},           % Strings in purple
    showstringspaces=false,                   % Do not show spaces in strings
    breaklines=true,                          % Automatic line breaking
    tabsize=4,                                % Set tab size
    numbers=none,                             % No line numbers
    frame=single, % No frame around code
    backgroundcolor=\color{backcolour},
    captionpos=b,                             % Caption position at bottom
    morekeywords={self, __init__, __name__, __main__}, % Additional keywords
}

\definecolor{codegray}{gray}{0.95}

\definecolor{baselinecolor}{gray}{.9}

\newcommand{\paragrapht}[1]{\noindent\textbf{#1}}  % tidy \paragraph

\usepackage{pifont}
\usepackage{amssymb}

\usepackage{xspace}
\usepackage{array}
\usepackage{layouts}
\usepackage{algorithm}
\usepackage{algpseudocode}
\usepackage{bbding}
\usepackage{listings}
\usepackage{pifont}
\usepackage{wasysym}

\usepackage{footnote}
\makesavenoteenv{tabular}
\makesavenoteenv{table}
\makesavenoteenv{figure}
\usepackage{pythonhighlight}

\algtext*{EndWhile}

\newcommand{\ie}{\textit{i.e.\xspace}}
\newcommand{\eg}{\textit{e.g.\xspace}}

\usepackage{xcolor}
\definecolor{mygreen}{RGB}{65,159,102}
\definecolor{myred}{RGB}{200,86,88}
\definecolor{dark_green}{rgb}{0, 0.5, 0}

\usepackage[most]{tcolorbox}
\newtcolorbox{promptbox}{colback=gray!5, colframe=gray!40, boxrule=0.4pt, arc=1pt, left=4pt, right=4pt, top=3pt, bottom=3pt, fontupper=\small\ttfamily}

\newcommand{\ours}{\textsc{AudioHelix}\xspace}
\newcommand{\graph}{{DualGraph}\xspace}

\newcommand{\onenode}{\mathbf{v}}
\newcommand{\chunknode}{\onenode_c}
\newcommand{\unitnode}{\onenode_u}
\newcommand{\conceptnode}{\onenode_s}
\newcommand{\bq}{\mathbf{q}}

\begin{document}
%\ninept
%
\maketitle

\begin{abstract}
Large audio language models (LALMs) struggle with long-form audio reasoning, as processing entire recordings often limits to capture query-relevant content distributed across distant segments while incurring substantial computational cost.
We introduce \ours, a hierarchical \graph framework tailored for LALMs' retrieval-augmented generation (RAG) systems.
Unlike existing text-only or non-graph RAG approaches, \ours connects related textual and acoustic evidence across temporally distant segments for multi-hop reasoning.
Extensive experiments on TraceAV-Bench, MMAU-Pro, and AudioMarathon demonstrate that \ours consistently outperforms full-audio processing and existing RAG methods.
The project site is \url{https://github.com/ewha-mmai/audiohelix}.
% Code is available on GitHub.\textsuperscript{1}

\end{abstract}


\section{Introduction}

A person who has heard a two-hour interview can answer questions about it without listening to the entire recording again. 
Instead, they recall or briefly revisit only the gold question-relevant moments to arrive at an answer.
What they recall contains not only linguistic information, but also rich paralinguistic cues, such as emotion, tone, and speaking style, that provide additional context for understanding the content.

LALMs~\cite{chu2024qwen2, ghosh2026audio,liu2025voxtral}, however, do not naturally have this ability when processing long-form audio. 
They typically receive the entire recording, processing a large amount of audio context even when only a small portion is relevant to the question.
Therefore, LALMs have struggled to identify where to focus while also incurring substantial computational and memory costs as the input grows longer~\cite{he2025audiomarathon}.
For example, a 35m recording occupies roughly 52.5K tokens~\cite{Qwen2.5-Omni}, far exceeding the number of tokens needed to represent its text transcript.
% requiring the LALM to process and reason over the entire input within its context window. % token 수는 Qwen2.5-Omni 기준

Long-form audio understanding therefore calls for retrieval: reasoning over only the query-relevant segments, not the full recordings, which is the principle underlying RAG~\cite{lewis2020retrieval}.
Text-only RAG~\cite{choudhari2025beamforming,jacobs2024leveraging, kang2025getalp}, which retrieves over ASR transcripts (or along with transcriptions), discards paralinguistic cues such as tone, pauses, and speaking style that carry information beyond the words themselves.
One might attempt to recover these cues by verbalizing them into text, for example through audio captioning.
Such descriptions, however, are prone to hallucination~\cite{kuan2024understanding}, and they are bounded by what the captioning prompt specifies.
On the other hand, audio-only RAG~\cite{mundnich2025speech,chen2025wavrag} is limited by the substantial redundancy in raw audio itself (\eg, silences, repeated phrasing) that introduces noise into the retrieval process.

\begin{figure}[!t]
    \centering
    \includegraphics[width=\linewidth]{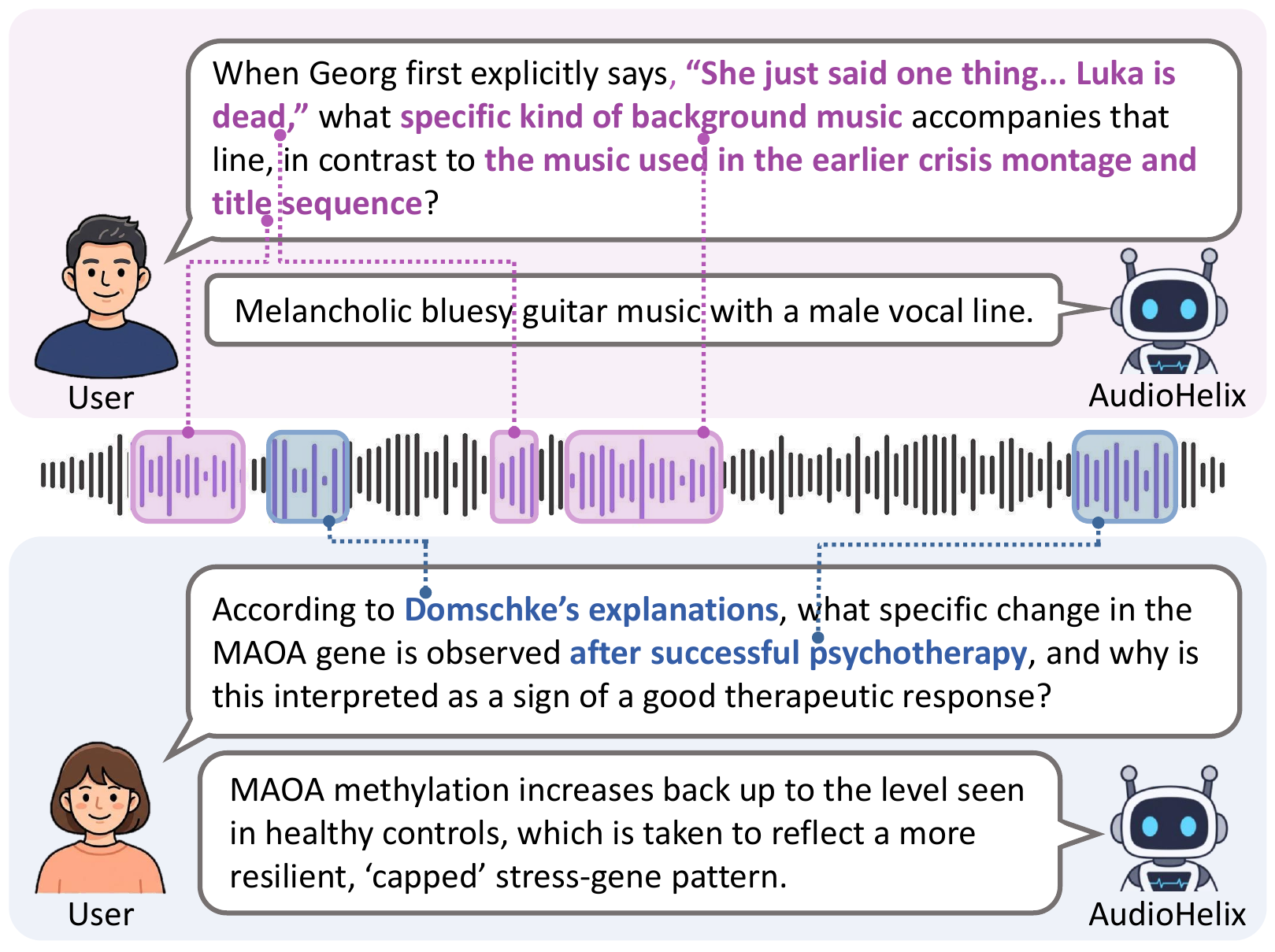}
    \vspace{-2em}
    \caption{{Long-form audio reasoning} requires concentrating on multi-hops to answer each query.}
    \label{figure1}\vspace{-1em}
\end{figure}

\begin{figure*}[!t]
    \centering
    \includegraphics[width=\linewidth]{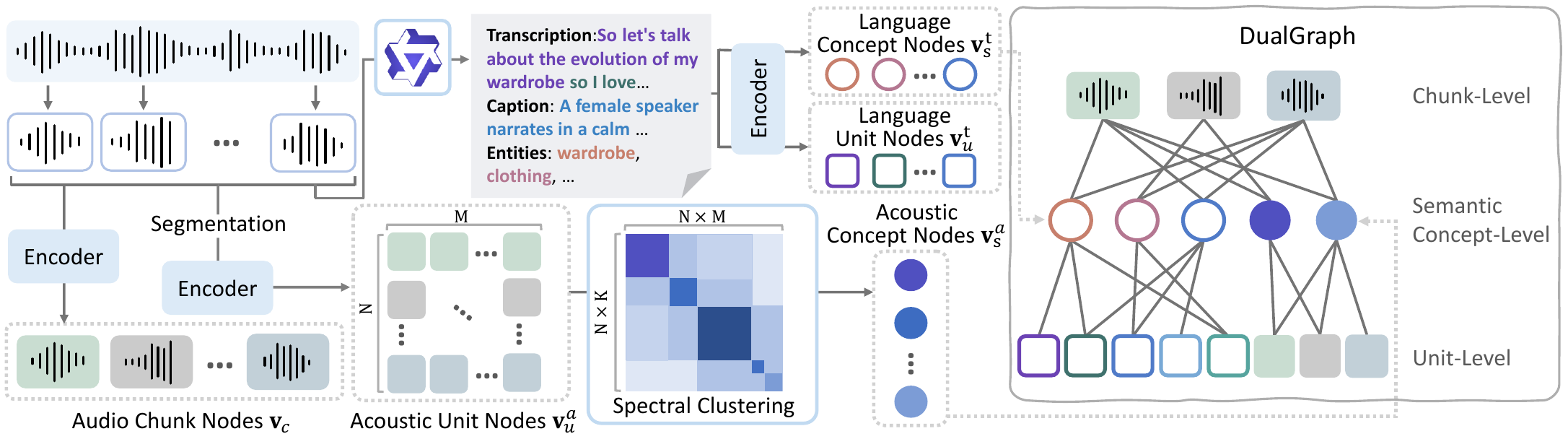}
    \vspace{-2em}
    \caption{The pipeline of building \graph in \ours. In the offline stage, the long-form audio is divided into fixed-length chunks, from which hierarchical fine-grained language and acoustic information is extracted to construct \graph with three levels: chunk-level, unit-level, and semantic concept-level.}
    \label{fig:framework}
    \vspace{-.5em}
\end{figure*}

In this paper, we propose \ours, a hierarchical \graph approach on RAG for long-form audio reasoning with LALMs. 
In contrast to text-only RAG for LALMs~\cite{someki2026planrag,hegde2026event}, \ours utilizes both acoustic and textual representations to build a hierarchical structure with three levels, including chunk-, unit-, and semantic concept-level, independent of any query.
% \ours extracts complementary textual and acoustic representations from each audio chunk and organizes them into \graph, a multimodal graph that connects related chunks through entity nodes. 
Given a query, \ours first identifies seed concept nodes, and propagates query-relevant scores to retrieve a small set of query-specific audio chunks.
% Across all subsequent queries on that recording, we reuse the graph to score cross-modal relevance scores for each audio chunk, and propagate them over the entity-based graph, allowing related audio chunks to reinforce one another. 
Therefore, we surface complementary evidence beyond isolated chunk-level matching~\cite{hegde2026event, someki2026planrag}.
It preserves fine-grained paralinguistic information while addressing the modality gap problem~\cite{liang2022mind}, which causes na\"ive cross-modal retrieval to underperform. 
By retrieving a compact set of related evidence, \ours provides downstream language models with informative context for reasoning over long-form audio, even as recordings extend beyond an hour.
To sum up, our contributions are three-fold:\vspace{-.7em}
\begin{itemize}
    \item To the best of our knowledge, \ours is a first attempt to build a hierarchical multimodal graph (\graph) for RAG with LALMs. \vspace{-.7em}
    % We propose \ours, a graph-based RAG framework for long-form audio reasoning that jointly leverages textual and acoustic information for fine-grained evidence retrieval.
    \item Our \graph connects audio chunks through shared concept nodes, enabling relevance propagation across related but temporally distant evidence. \vspace{-.7em}
    \item Extensive experiments demonstrate the effectiveness of \ours with three popular LALMs for long-form audio understanding on TraceAV-Bench, MMAU-Pro, and AudioMarathon.
    \vspace{-1em}
\end{itemize}

\section{Method}
\subsection{Motivation and Overview}\vspace{-1pt}
Long-form audio reasoning requires LALMs to concentrate on the query-relevant portion of an input stream~\cite{feng2026traceav}, since passing the entire track exceeds its effective context and computation budget~\cite{he2025audiomarathon, chaichana2026extending,wang2025timeaudio}.
Prior LALM-RAG methods~\cite{ penamakuri2025audiopedia, hegde2026event,someki2026planrag} have relied on text to explicitly represent semantic cues.
Representing audio purely through text, however, often fails to ground acoustic detail.
% Recent RAG approaches for long-form understanding commonly rely on textual representations~\cite{xue2025adavideorag, ventura2025chapter, luo2026video,hegde2026event, someki2026planrag}. These representations provide explicit semantic cues and facilitate the organization and retrieval of conceptually related audio evidence. However, textual representations may lose fine-grained acoustic information that is difficult to describe comprehensively, including paralinguistic cues and subtle acoustic events~\cite{zhang2026beyond}. 
Conversely, directly using acoustic representations may struggle to associate acoustically distinct instances of similar semantic concepts~\cite{araz2024evaluation}.
% Accordingly, textual and acoustic representations capture different aspects of audio evidence. \ours brings them together to enable retrieval based on both semantic relatedness and fine-grained acoustic characteristics.

Our \ours constructs a multimodal graph from both textual and acoustic representations.
Rather than explicitly learning unified cross-modal embedding~\cite{mundnich2025speech, chen2025wavrag}, we preserve their own features with hierarchical structure, then link them at the semantic concept level.
Specifically, given a long audio track $A$, \ours generates a \graph $\mathcal{G}=(\mathcal{V}, \mathcal{E})$, where each node $\onenode\in\mathcal{V}$ represents text or audio embeddings derived from $A$, and each edge in $\mathcal{E}$ encodes a relation between them, illustrated in \cref{fig:framework}.
We carefully note that the graph is constructed once per recording in an offline stage, following prior video-tailored approaches~\cite{shen2026vgent,jin2026coarse, jeong2025videorag}.
To answer the given text query $Q$, \ours retrieves an audio subset $A_{u^*}$ by traversing the graph.
% Only the audio segments associated with $\mathcal{V}_{q}$ are passed to the LALM to produce the answer.
The LALM thus reasons over a small, query-specific audio segment and related text evidence, rather than the entire recording.

\subsection{\graph}\vspace{-3pt}
% Given a long-form audio recording $A$, we construct a hierarchical dual graph, termed \graph, that consists of textual and acoustic nodes.
\ours first segments audio input $A$ into fixed-length audio chunks $\{c_i\}_{i=1}^{N}$, from which modality-specific evidence is subsequently extracted.
However, a single level of granularity is insufficient to serve queries that require different degrees of context.
Coarse segments preserve context but include irrelevant audio, while concepts are precise but lack context.
To adaptively match the granularity a query demands, our \graph is structured into three hierarchical levels, namely chunk-, unit-, and semantic concept-level.

% \paragrapht{Chunk-level.} 

% Each node corresponds to an audio chunk $c_i$ and represented by an acoustic embedding of the chunk at the coarsest granularity.
% We additionally extract an ASR transcription and event description for each chunk using a LALM~\cite{Qwen2.5-Omni}, as shown in \cref{fig:framework}.
% This textual information serves as auxiliary context rather than a retrieval representation.
% \todo{audio chunk node...}

\paragrapht{Chunk- and Unit-level.}
The chunk level comprises audio chunk nodes $\chunknode$.
For unit-level nodes, we first extract textual content from each chunk with Qwen2.5-Omni~\cite{Qwen2.5-Omni}, obtaining a textual description including ASR and event caption. 
Each chunk is then divided into a set of $M$ fixed-length acoustic units, each represented by an acoustic embedding from pretrained audio-text encoders~\cite{vyas2026pushing} to form the acoustic unit nodes $\unitnode^a$.
% The extracted text is likewise decomposed into sentence-level segments, and each segment is embedded to form the language unit nodes $\unitnode^t$.
The textual information extracted from each chunk is likewise segmented into sentences, with each sentence embedded to form a language unit node $\unitnode^t$.

% We , and unit node consists of fine-grained language and acoustic unit nodes $\unitnode=\unitnode^T\cup\unitnode^A$, corresponding to sentence-level linguistic information and localized acoustic patterns, respectively.
% Finally, the entity level consists of language and acoustic entity nodes $\mathcal{V}_E=\mathcal{V}_E^T\cup\mathcal{V}_E^A$ which serves as anchors for bridging the unit and chunk levels.
% The resulting graph is defined as:
% \begin{equation}
%     \mathcal{G}=(\mathcal{V}, \mathcal{E}),\quad \mathcal{V}=\mathcal{V}_C\cup\mathcal{V}_U\cup\mathcal{V}_E.
% \end{equation}

% All graph nodes are represented in a shared audio-text embedding space using a pretrained audio-text encoder~\cite{vyas2026pushing}, facilitating the association of language and acoustic information within the graph.

\paragrapht{Semantic concept-level.}
While chunk-level and unit-level acoustic nodes are obtained directly by segmenting the audio, concept-level nodes are extracted separately to link related evidence across chunks and units.
For acoustic concepts, we group acoustically similar unit embeddings across all chunks using spectral clustering~\cite{ng2001spectral}, and take each cluster centroid as an acoustic concept node $\conceptnode^a$.
Meanwhile, the LALM also extracts a set of textual concepts mentioned in each chunk, which form the language concept nodes $\conceptnode^t$.

% \todo{edge}
\paragrapht{Edge.}
\graph establishes edges based on hierarchical inclusion.
Specifically, each concept node is connected to the unit nodes from which it is derived.
% It also connects to their parent chunk nodes, weighted by the proportion of the chunk's units connected to the concept.
It also connects to their parent chunk nodes, weighted by the proportion of the chunk's connections to the concept.
% thereby linking chunk-level and unit-level evidence through semantic concepts.
% For the acoustic modality, however, a single cluster assignment provide only a limited view of the relationships between the nodes.
% We further compensate acoustic similarity by linking each acoustic unit to its nearest $n$ acoustic concept nodes based on embedding similarity.
We further compensate acoustic similarity by linking each acoustic unit to its nearest $n$ acoustic concept nodes based on embedding similarity, retaining those whose similarity exceeds a threshold $\gamma$ and assigning edge weights in the same manner as above.

% \paragrapht{Language node.}
% To extract linguistic information from each audio chunk $c_i$, we process the chunk with a LALM~\cite{Qwen2.5-Omni}, obtaining a transcription, an audio caption, and textual entities.
% The transcription and caption are further decomposed into sentence-level units, which constitute the language unit nodes $\mathcal{V}_U^T$, while the extracted entities are represented as language entity nodes $\mathcal{V}_E^T$.
% The language entities represent semantic concepts mentioned in the transcription or caption.
% Each language entity node is connected to the language unit nodes in which it appears and the audio chunk node from which the corresponding language units are extracted, serving as an anchor that bridges the chunk and unit levels.

% \paragrapht{Acoustic node.}
% To obtain fine-grained acoustic information from each audio chunk
% $c_i$, we divide it into fixed-length acoustic units.
% These unit form the acoustic unit nodes $\mathcal{V}_U^A$, capturing acoustic evidence that may not be explicitly reflected in the textual representations.
% Based on their acoustic representations, the units are clustered using spectral clustering, with each cluster centroid forming an acoustic entity node in $\mathcal{V}_E^A$.
% Each acoustic unit node is connected to its $n$ nearest acoustic entity nodes, allowing it to capture multiple related acoustic patterns.
% The acoustic entity nodes are further connected to the audio chunk nodes from which their corresponding acoustic units are extracted, linking fine-grained acoustic patterns to their source chunks.

\subsection{Retrieval-Augmented Generation}\vspace{-1pt}
Given a query text embedding $\bq$ from~\cite{vyas2026pushing}, our goal is to retrieve a query-relevant node set $\{\onenode_{q}\}$.
% We first retrieve relevant audio chunks through relevance score propagation across concepts, retaining their textual information.
% For each retrieved chunk, we additionally select query-relevant acoustic units to provide complementary acoustic information.
Specifically, \ours starts retrieval from each modality's concept node most similar to the query, $\{\conceptnode^{t,n}, \conceptnode^{a,m}\}$.
For brevity, we omit the modality superscripts $t,a$ hereafter.
Their node weights $\{R_s^{n}, R_s^{m}\}$ are initialized with the corresponding similarities, respectively.
To propagate relevance between concept nodes through their shared units, for each unit connected to both $\conceptnode^i$ and a neighboring concept $\conceptnode^j$, we update the concept weight by query-unit similarity:
\begin{equation}
\label{eq:1}
R(\conceptnode^j)\leftarrow R(\conceptnode^j)+\sum\nolimits_{\unitnode \in \mathcal{N}(\conceptnode^i,\conceptnode^j) }R(\conceptnode^i)\cdot\operatorname{sim}(\bq, \unitnode^i),
\end{equation}
where $\operatorname{sim}(\cdot)$ denotes the cosine similarity.
The same process is then repeated until the maximum propagation step $P$.
Note that a unit node already visited during propagation is not revisited in later steps.

Motivated by \cite{zhuang2026linearrag}, we aggregate concept scores into a chunk, weighted by a coefficient $W_c$=$0.05$ and edge $\mathcal{E}$,
\begin{equation}
    R(\chunknode^i) = W_c \cdot \psi( 
    \sum\nolimits_{\conceptnode^j \in \mathcal{N}(\chunknode^i)}  \frac{R(\conceptnode^j) \cdot\psi(\mathcal{E}(\conceptnode^j, \chunknode^i))}{p^j} 
    ),
    \label{eq:2}
\end{equation}
where $p^j$ denotes the propagation step at which $\onenode^j$ first receives a nonzero score in \cref{eq:1}, and $\psi(x)=\ln(1+x)$ following  retrieval scoring~\cite{zhuang2026linearrag}.
We then propagate $R(\chunknode)$ across the graph via Personalized PageRank (PPR)~\cite{haveliwala2003analytical,zhuang2026linearrag}.
% For an audio chunk $c_i$, we compute its chunk relevance score as:
% \begin{equation}
%     r_C(c_i)=W_C\ln\big(1+\sum_{e_j}r_E(e_j)\frac{\ln(1+M_C(c_i,e_j))}{L_{e_j}})\big),
% \end{equation}
% where $r_E(e_j)$ denotes the relevance score of entity $e_j$, $M_C(c_i,e_j)$ represents the association weight between the chunk and entity, $L_{e_j}$ denotes the propagation depth of the entity, and $W_C$ is a scaling factor.
% We then perform Personalized PageRank over the graph using the resulting entity and chunk relevance scores to compute a global importance score for each node.

% Algorithm formatting
\algrenewcommand\algorithmiccomment[1]{%
    \hfill\textcolor{gray}{$\triangleright$ #1}}
\definecolor{commentpurple}{rgb}{0.45, 0.35, 0.6}
\newcommand{\purple}[1]{\textcolor{commentpurple}{#1}}
\algrenewcommand\algorithmicrequire{\textbf{Input:}}
\algrenewcommand\algorithmicensure{\textbf{Output:}}
\algnewcommand\algorithmicinitialize{\textbf{Initialize:}}
\algnewcommand{\Initialize}{\item[\algorithmicinitialize]}

\algnewcommand\algorithmicparameter{\textbf{Parameter:}}
\algnewcommand{\Parameter}{\item[\algorithmicparameter]}

\begin{algorithm}[t]
\caption{RAG in \ours}
\label{alg:retrieval}
\begin{algorithmic}[1]

\Require $A$, $D$, $Q$, $\bq$
\Ensure $Y$

\Parameter $P$, $K_u$, $K_c$

\Initialize $R(\conceptnode) \leftarrow 0$, $\conceptnode^0 \leftarrow \operatorname{argmax}(\bq, \{\conceptnode\})$

% \Statex \purple{// 1. Concept Relevance Propagation}
% \State $\mathcal{S}\leftarrow
% \big\{\operatorname*{arg\,max}_{s\in\mathcal{V}_s^t}\operatorname{sim}(Q,s),\;
% \operatorname*{arg\,max}_{s\in\mathcal{V}_s^a}\operatorname{sim}(Q,s)\big\}$

% \State $\mathcal{S}\leftarrow
% \{\conceptnode^{t,n},\conceptnode^{a,m}\}$

% \State $R(\conceptnode)\leftarrow 0,\quad \forall\conceptnode\in\mathcal{V}_s$

\State $R(\conceptnode^0)\leftarrow\operatorname{sim}(\bq,\conceptnode^0)$

% \State
% \State for $p < P$\todo{..}
\While {$p < P$}
    \State $R(\onenode^{p+1})$ 
    $\leftarrow$ Propagation($R(\onenode^p)$) \Comment{\Cref{eq:1}, \cref{eq:2}}
    % \State $\displaystyle
    % R(\conceptnode^j)\leftarrow R(\conceptnode^j)+
    % \sum_{\conceptnode^i\in\mathcal{S}}
    % \sum_{\unitnode\in\mathcal{N}_{ij}}
    % R(\conceptnode^i)\operatorname{sim}(Q,\unitnode)$
    
    % \State $\mathcal{S}\leftarrow
    % \{\conceptnode^j\mid
    % \exists\conceptnode^i\in\mathcal{S},\
    % \mathcal{N}_{ij}\neq\emptyset\}$
    
    % \Comment{$\mathcal{N}_{ij}=
    % \mathcal{N}(\conceptnode^i)\cap\mathcal{N}(\conceptnode^j)$}
\EndWhile
% \algtext*{EndWhile}

% \Statex \purple{// 2. Audio Chunk Retrieval}
% \State $R(\chunknode^i) = W_c \cdot \psi( 
    % \sum\nolimits_{\conceptnode^j \in \mathcal{N}(\chunknode^i)}\frac{R(\conceptnode^j) \cdot\psi(\mathcal{E}(\conceptnode^j, \chunknode^i))}{L_j} 
    % )$
% \State $\{ \chunknode^* , \onenode_q \}\leftarrow
% \operatorname{TopK_c}((\operatorname{PPR}\!\left(\mathcal{G}\right)))$, $\operatorname{TopK_u}\{\operatorname{sim}(\bq,\onenode_{u \in c^*})\}$

\State $\{ c^*  \}\leftarrow
\operatorname{Top-K_c}(\operatorname{PPR}\!\left(\mathcal{G}, R(\chunknode)\right))$

% \State $\{ u^* \}\leftarrow
%  \operatorname{Top-K_u}\{\operatorname{sim}(\bq,\{ \onenode_{c^*}  \})\}$

\State $\{u^*\}\leftarrow
\operatorname{Top-K_u}\{\operatorname{sim}(\bq,\onenode_u^a):u\in c^*\}$

% \State
% \Statex \purple{// 3. Audio Unit Retrieval}
% \State $\{\onenode_q\}\leftarrow
% \operatorname{TopQ}\{\operatorname{sim}(Q,\onenode_{u \in c^*})\}$

% \State
% \Statex \purple{// 4. Answer Generation}
% \State $Y \leftarrow\operatorname{LALM}\!\left([A_{\{\onenode_q\}}^{c_1};
% D_{\{\onenode_q\}}^{c_1};
% \cdots;
% A_{\{\onenode_q\}}^{c_K};
% D_{\{\onenode_q\}}^{c_K};
% Q]\right)$
% \State \Return $\hat{y}$
\State $Y\leftarrow\operatorname{LALM}([\{A_{u^*};D_{c^*}\};Q])$

\end{algorithmic}
\end{algorithm}

The $\operatorname{Top-K_c}$ audio chunks are retrieved by their global relevance scores $R(\chunknode)$, for which $\operatorname{Top-K_u}$ acoustic units $A_{u^*}$ are selected according to their relevance to the query, further reducing the inference cost.
For each retrieved chunk, we include its transcription and caption $D_{c^*}$, together.
% Therefore, we concatenate audio unit-level segments, textual description, and instruction into a unified prompt $[{A}_{\{\onenode_q\}};D_{\{\onenode_q\}};Q]$, assigning each a positional embedding based on its original timestamp in $A$.
We then organize the selected audio units, textual descriptions in their original temporal order, forming a chunk-wise input sequence $[\{A_{u^*};D_{c^*}\};Q]$ to generate a final answer $Y$.
Since Audio-Flamingo-3~\cite{ghosh2026audio} does not support multiple inputs, audio units are unified into a single recording with positional embeddings for their original timestamps.
% LALMs then generate a final answer $Y$.
The overall RAG in \ours is demonstrated in \cref{alg:retrieval}.

\section{Experiments}
\begin{table}[t]
\centering
\small
\setlength{\tabcolsep}{3pt}

% \resizebox{\columnwidth}{!}{%
\begin{tabular}{@{}lccc@{}}
\toprule
Method & TraceAV & MMAU-Pro & AM \\
\midrule

Audio-Flamingo-3~\cite{ghosh2026audio}
& \underline{56.9} & \underline{50.7} & \underline{57.3} \\

\quad + LinearRAG~\cite{zhuang2026linearrag} 
%& 54.4 & 47.9 & 56.1 \\
& 54.4 & 48.0 & 56.1 \\

\quad + PlanRAG-Audio~\cite{someki2026planrag} 
& 37.5 & 34.0 & 30.8 \\

\rowcolor{blue!5}\quad + \ours
% & \textbf{59.7} & \textbf{52.6} & \textbf{58.4} \\
& \textbf{59.4} & \textbf{52.3} & \textbf{58.5} \\
\midrule

Voxtral-3B~\cite{liu2025voxtral}
& \underline{48.1} & 40.9 & \underline{56.4} \\

\quad + LinearRAG~\cite{zhuang2026linearrag} 
%& 38.4 & \underline{53.1} & 52.3 \\
& 38.9 & \underline{53.1} & 52.3 \\

\quad + PlanRAG-Audio~\cite{someki2026planrag} 
& 44.6 & 35.4 & 39.0 \\

\rowcolor{blue!5}\quad + \ours
% & \textbf{49.3} & \textbf{56.8} & \textbf{57.7} \\
& \textbf{50.6} & \textbf{56.5} & \textbf{57.9} \\

\midrule

Qwen3-Omni-30B~\cite{xu2025qwen3}
& \underline{71.3} & \underline{66.1} & \underline{63.7} \\

\quad + LinearRAG~\cite{zhuang2026linearrag} 
%& 62.9 & 60.5 & 61.5 \\
& 62.2 & 61.5 & 61.9 \\

\quad + PlanRAG-Audio~\cite{someki2026planrag} 
& 55.3 & 34.4 & 37.0 \\

\rowcolor{blue!5}\quad + \ours
% & \textbf{72.3} & \textbf{66.9} & \textbf{66.4} \\
& \textbf{71.6} & \textbf{67.2} & \textbf{66.1} \\

\bottomrule
\end{tabular}%
% }
\vspace{-.7em}
\caption{Performance comparison on three benchmarks. Metrics are accuracy for TraceAV-Bench and MMAU-Pro, and F1 score for AudioMarathon (AM).
The \textbf{best} score is bold, and the \underline{secondary} is underlined.}
\label{tab:main}
\vspace{-.5em}
\end{table}
\subsection{Experimental Settings}
\paragrapht{Benchmarks \& baselines. }
We evaluate performance on three long audio benchmarks; 
% TraceAV-Bench evaluates multi-hop reasoning over long audio-visual recordings.
% For audio-based evaluation, we use its audio-centric reasoning (AR) dimension, which covers speech context (\textbf{SC}), environmental sound (\textbf{ES}), and background music (\textbf{BM}).
On TraceAV-Bench~\cite{feng2026traceav}, we assess multi-hop reasoning using its audio-centric reasoning (AR) subset, with publicly downloadable videos.
For MMAU-Pro~\cite{kumar2026mmau}, we use the long and ultra-long subsets, covering $<$10m recordings.
For AudioMarathon~\cite{he2025audiomarathon}, we use all tasks except ASR, since it requires full transcription rather than reasoning.
We use three representative LALMs, including Audio-Flamingo(AF)-3~\cite{ghosh2026audio}, Voxtral-3B~\cite{liu2025voxtral}, and Qwen3-Omni-30B~\cite{xu2025qwen3}.
We compare our \ours with LinearRAG~\cite{zhuang2026linearrag} and PlanRAG-Audio~\cite{someki2026planrag}.

% \input{table/linaerrag_comp}

% \input{table/topk}
% \begin{figure}[!t]
%     \centering
%     \includegraphics[width=0.8\linewidth]{fig/figure3.pdf}
%     \vspace{-1em}
%     \caption{\textbf{Effect of top-$k$ audio chunk retrieval} on TraceAV-Bench. The dashed line denotes the baseline performance.}
%     \label{fig:topk_ablation}
%     \vspace{-1em}
% \end{figure}

% For LinearRAG, we construct a text-only graph from descriptions extracted from the audio and provide the retrieved text to the base models for answer generation.
% For PlanRAG-Audio, an RAG method designed for long-form audio understanding, we reimplement the method based on the original paper, as its implementation is not publicly available.

\paragrapht{Implementation details. }
We use 45s chunks and 5s acoustic units for TraceAV-Bench, and 30s chunks and 6s units for more shortly MMAU-Pro and AudioMarathon.
% For each chunk, Qwen2.5-Omni~\cite{Qwen2.5-Omni} jointly extracts transcriptions, captions, and textual entities in a single inference, and all graph nodes are represented using the Perception Encoder AudioVisual~\cite{vyas2026pushing}.
Acoustic units connect to their top-3 nearest centroids of 10 clusters with $\gamma$= $0.85$.
We set $\operatorname{Top-K_c}$ to 15 for TraceAV-Bench and 8 for the remaining benchmarks, selecting two acoustic units per chunk (\ie, $K_u$=$2$).
Baseline methods follow their respective retrieval procedures with the same base models and answer generation settings.
Note that LinearRAG uses descriptions extracted from the entire recording to make the corpus, while PlanRAG-Audio follows the paper.

%\begin{table}[t]
%\centering
%\caption{Ablation study on TraceAV-Bench~\cite{feng2026traceav} using Audio Flamingo3~\cite{ghosh2026audio} / Qwen3-Omni-%30B~\cite{xu2025qwen3} / Voxtral-Mini-3B-2507~\cite{liu2025voxtral} as the base model.}
%\label{tab:ablation}
%\small
%\begin{tabularx}{\columnwidth}{l *{4}{>{\centering\arraybackslash}X}}
%\toprule
%Method & SC & ES & BM & Avg. \\
%\midrule
%Audio-Flamingo-3 &66.3 &59.2 &46.7 & 56.9 \\
%+ Oracle & 69.9 & 58.0 & 55.6 & 61.3 \\
%\midrule
%+ Na\"iveRAG & 61.0& 54.3 &58.0 &58.1  \\
%+ Ours w/o Audio & 64.6 & 56.7 & 55.6 & 59.1 \\
%+ Ours w/o Text & 59.2 & 51.8 & 57.2 & 56.6 \\
%\rowcolor{gray!10}
%\textbf{+ Ours} & 63.7 & 59.2 & 56.4 & 59.7 \\

%\midrule
%Voxtral-Mini-3B-2507 &62.8 & 41.9& 38.7& 48.1  \\
%+ Oracle &- & -& - & -\\
%\midrule
%+ Na\"iveRAG & - &- & - & - \\
%+ Ours w/o Audio & - & - &- &- \\
%+ Ours w/o Text & -& - & - &- \\
%\rowcolor{gray!10}
%\textbf{+ Ours}  & 59.2 & 43.2 & 44.3 & 49.3 \\

%\midrule
%Qwen3-Omni-30B &87.6 &62.9 &62.0 &71.3 \\
%+ Oracle & 92.0 & 63.0 & 66.1 & 67.9 \\
%\midrule
%+ Na\"iveRAG & 66.3 & 55.5 & 70.1 & 65.0 \\
%+ Ours w/o Audio & 76.9 & 65.4 & 58.8 & 66.9 \\
%+ Ours w/o Text & 66.3 & 56.7 & 70.1 & 65.4 \\
%\rowcolor{gray!10}
%\textbf{+ Ours} & 80.5 & 65.4 & 69.3 & 72.3 \\

%\bottomrule
%\end{tabularx}
%\end{table}

\begin{table}[t]
\centering
\small
\setlength{\tabcolsep}{2pt}

% \resizebox{\columnwidth}{!}{%
\begin{tabular}{@{}lccc@{}}
\toprule
Method & TraceAV & MMAU-Pro & AM \\
\midrule

Audio-Flamingo-3 &56.9 & 50.7&57.3  \\
% \quad + Oracle & 61.3 & -&- \\
% \midrule
%\quad + Ours w/o Text &56.6 &47.9 &55.2 \\
\quad + \ours (w/o Text) &55.3 & 48.2&54.4 \\
% 비교용 \quad + LinearRAG   & 54.4 &  48.0 & 56.1 \\ 
\quad + \ours (w/o Audio) &58.8 & 50.7&56.5 \\

\quad + Na\"iveRAG & 58.1 &58.0 &62.6  \\
\quad + LinearRAG (w/ Our text)   & 60.0 &  50.0 & 54.7 \\ 

% \rowcolor{gray!10}
\rowcolor{blue!5}
% \quad \textbf{+ Ours} &59.7 &52.6 &58.4 \\
\quad \textbf{+ \ours} & 59.4 & 52.3 & 58.5 \\

\midrule
Voxtral-3B &48.1 &40.9 &56.4 \\
% \quad + Oracle &46.9 & -& - \\
% \midrule
%\quad + Ours w/o Text &42.5 &47.1 &46.1  \\
\quad + \ours (w/o Text) &41.1 &46.7 & 45.5 \\
\quad + \ours (w/o Audio) & 44.9 &53.1 &58.1 \\

\quad + Na\"iveRAG & 42.7 &30.5 &52.8 \\
\quad + LinearRAG (w/ Our text)   &45.9& 54.9& 54.8 \\
%비교용 \quad + LinearRAG    &38.9& 53.1& 52.3 \\

% \rowcolor{gray!10}
\rowcolor{blue!5}
% \quad \textbf{+ Ours}  &49.3 &56.8 &57.7 \\
\quad \textbf{+ \ours} & 50.6 & 56.5 & 57.9 \\

\midrule
Qwen3-Omni-30B &71.3 &66.1 &63.7 \\
%\quad + Ours w/o Text &65.4 &64.5 &63.3 \\
\quad + \ours (w/o Text) &67.2 &64.4 &64.2\\
% \quad + Oracle &67.9& -& -\\
% \midrule
\quad + \ours (w/o Audio) &65.7 &60.5 &64.4 \\

\quad + Na\"iveRAG & 65.0 &66.4 &62.9 \\
\quad + LinearRAG (w/ Our text)   & 66.9 &62.8&  62.6 \\
%비교용 \quad + LinearRAG   &62.2& 61.5& 61.9 \\

% \rowcolor{gray!10}
\rowcolor{blue!5}
% \quad \textbf{+ Ours} &72.3 &66.9 &66.4 \\
\quad + \textbf{\ours} & 71.6 & 67.2 & 66.1 \\

\bottomrule

\end{tabular}%
\vspace{-.5em}
\caption{Ablation study on graph-based retrieval, graph construction, and modality contributions.}
\label{tab:ablation}
% }
\vspace{-.5em}
\end{table}

% \begin{table}[t]
% \centering
% \caption{Ablation study on TraceAV-Bench~\cite{feng2026traceav} using Qwen2.5-Omni-7B~\cite{Qwen2.5-Omni} as the base model.}
% \label{tab:ablation}
% \small
% \begin{tabularx}{\columnwidth}{l *{4}{>{\centering\arraybackslash}X}}
% \toprule
% Method & SC & ES & BM & Avg. \\
% \midrule
% Qwen2.5-Omni-7B & 67.2 & 46.9 & 56.4 & 57.8 \\
% + Oracle & 77.8 & 51.8 & 58.0 & 63.5 \\
% \midrule
% + Na\"iveRAG & - & - & - & - \\
% + Ours w/o Audio & 58.4 & 42.0 & 37.1 & 45.9 \\
% + Ours w/o Text & - & - & - & - \\
% \rowcolor{gray!10}
% \textbf{+ Ours} & 71.6 & 58.0 & 58.8 & 63.2 \\
% \bottomrule
% \end{tabularx}
% \end{table}

\subsection{Results}
\cref{tab:main} shows the performance comparsion on three benchmarks, showing that \ours achieves state-of-the-art.
While other methods often degrade performance, \ours consistently improves over the corresponding base LALMs despite using only a small subset of the original audio.
% As shown in \cref{tab:inference_cost}, \ours reduces latency by 11.1$\times$ and computation by 17.8$\times$.
% In contrast, the existing RAG baselines often degrade performance.
% For example, on MMAU-Pro, LinearRAG and PlanRAG-Audio decrease the performance of Qwen3-Omni by 5.6 and 31.7 points, respectively, whereas \ours improves it by 1.1 points.
% These results highlight the importance of retrieving only relevant evidence from long-form audio for multi-hop reasoning, rather than processing the entire recording as input.
Especially, \ours outperforms PlanRAG-Audio~\cite{someki2026planrag} by 6.0 to 32.8 across all LALMs and benchmarks, while \ours also obtains higher scores than LinearRAG~\cite{zhuang2026linearrag} on all benchmarks.
The results demonstrate that modeling multimodal relationships as a graph in \ours captures complementary evidence across temporally distinct chunks, while supporting fine-grained acoustic cues compared to text-only RAG~\cite{zhuang2026linearrag,someki2026planrag}.
Qualitative results of the retrieval and QA are provided at \url{http://mmai.ewha.ac.kr/audiohelix/}.

% Notably, \ours also obtains higher scores than LinearRAG on all benchmarks.
% While LinearRAG can effectively retrieve multi-hop textual evidence, its reliance on textual representations prevents it from capturing acoustic characteristics that may provide complementary evidence for audio questions.
% By incorporating both textual and acoustic representations, \ours can retrieve complementary information across modalities, resulting in improved performance over LinearRAG.

\subsection{Dicussion}
\paragrapht{Audio-text complementarity.}
Removing either modality consistently degrades performance, with the text-only variant retaining more of the full performance than the acoustic-only variant.
Therefore, acoustic evidence alone is less effective for relevance propagation, while combining both modalities yields consistent gains over either variant alone.

\paragrapht{Text parsing.}
We validate our text parsing and graph contribution by replacing LinearRAG's graph with our text-only graph, \ie, `LinearRAG (w/Our text)' in \cref{tab:ablation}.
This variant improves over the original LinearRAG's performance on \cref{tab:main}, proving that our \graph is more robust to retrieval.

\begin{figure}[t]
    \centering
    \begin{subfigure}{0.495\linewidth}
        \centering
        \includegraphics[width=\linewidth]{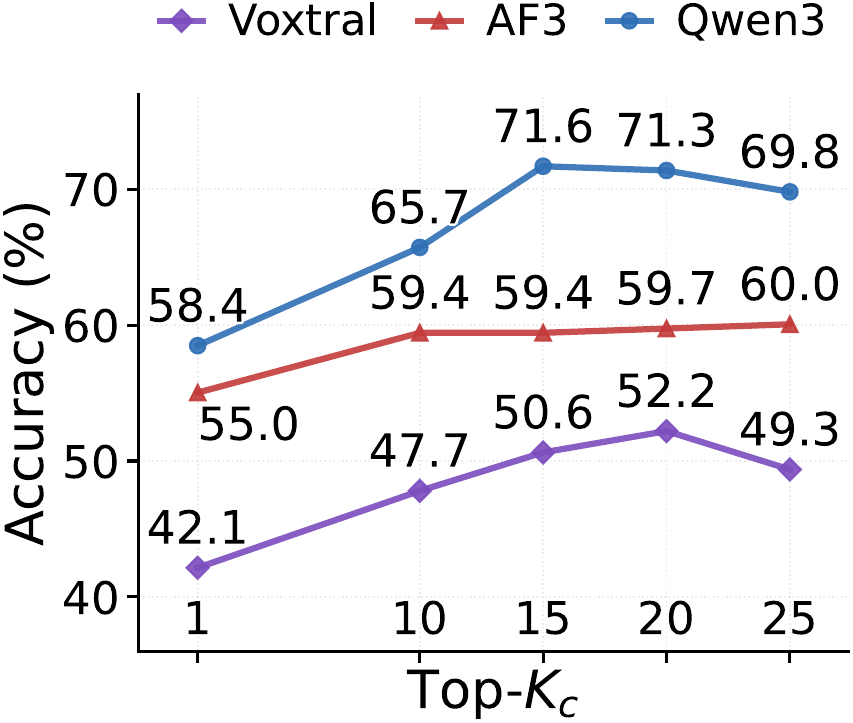}
        \caption{}
        \label{fig:figure3_a}
    \end{subfigure}
    \hfill
    \begin{subfigure}{0.495\linewidth}
        \centering
        \includegraphics[width=\linewidth]{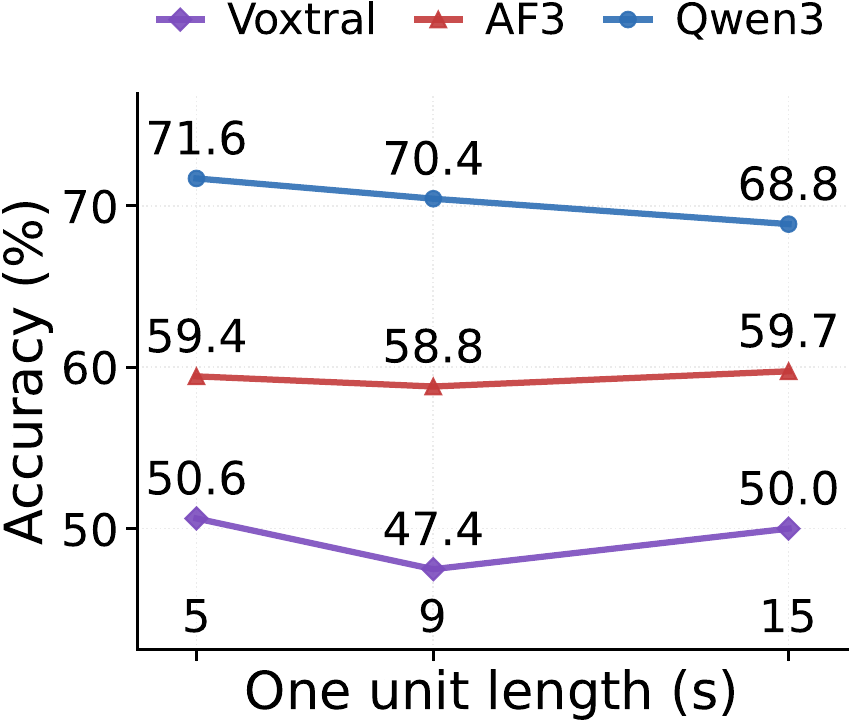}
        \caption{}
        \label{fig:figure3_b}
    \end{subfigure}
    \vspace{-2em}
    \caption{
Performance on TraceAV-Bench varying (a) number of retrieved audio chunks and
(b) acoustic unit lengths.}
    \label{fig:figure3}
    \vspace{-.5em}
\end{figure}

\paragrapht{Graph structure.}
As a non-graph baseline, we further compare \ours with Na\"iveRAG~\cite{lewis2020retrieval}, which ranks audio units directly by query relevance and retrieves the same total duration of audio.
The results in \cref{tab:ablation} indicate that modeling hierarchical associations surfaces indirectly connected units.
This gap is most pronounced for Voxtral-3B on MMAU-Pro, where \ours improves 15.6\% while Na\"iveRAG degrades it by 10.6\%.

\begin{table}[t]
\centering
\resizebox{\linewidth}{!}{
\begin{tabular}{lccc}
\toprule
Method & Indexing (s) $\downarrow$ & Latency (s) $\downarrow$ & TFLOPs $\downarrow$   \\
\midrule
Full Audio
%& - &8.36 &$20.61 \times 10^2$\\
& - &17.52 &$20.61 \times 10^2$\\
LinearRAG~\cite{zhuang2026linearrag}
%& 108.81 & 0.31&$0.42 \times 10^2$ \\
& 169.77 & 2.57 &$0.41 \times 10^2$ \\
PlanRAG-Audio~\cite{someki2026planrag}
%&62.67 &16.59 &$1.26 \times 10^2$ \\
&476.99 &18.18 &$2.28 \times 10^2$ \\
\ours
%&169.08 &0.75 &  $1.16 \times 10^2$ \\
&391.45 &1.58 &  $1.17 \times 10^2$ \\
\bottomrule
\end{tabular}
}
\vspace{-.7em}
\caption{Average inference cost of Audio Flamingo 3~\cite{ghosh2026audio} on 100 randomly sampled instances from TraceAV-Bench. 8$\times$RTX Pro 6000 are used.}\label{tab:inference_cost}
\vspace{-1em}
\end{table}

% To assess the effectiveness of graph-based retrieval and multimodal evidence, we compare \ours with Na\"iveRAG and variants that remove either the textual or acoustic modality from the graph.
% The results are shown in \cref{tab:ablation}.
% Graph-based retrieval consistently outperforms Na\"iveRAG, demonstrating the benefit of modeling relationships among audio segments for multi-hop reasoning.
% The limited gains of the single-modality variants suggest that each modality alone may not sufficiently capture the relationships, while their combination provides complementary evidence for multi-hop reasoning.
% The text-only variant generally outperforms Na\"iveRAG, suggesting that modeling hierarchical relationships among the evidence is beneficial.
% However, the acoustic-only variant generally underperforms Na\"iveRAG, indicating that acoustic evidence alone is less amenable to such relation modeling.
% The full multimodal graph consistently outperforms both single-modality variants, suggesting that textual and acoustic evidence provide complementary cues for multi-hop reasoning.
% To further examine our graph construction, \cref{tab:linearrag} compares LinearRAG with a variant using our \graph.
% While LinearRAG constructs its graph from textual information extracted from the entire audio recording, \graph extracts textual information separately from each audio chunk separately.
% This chunk-level extraction provides finer-grained textual information for graph construction, resulting in consistently higher performance across the three benchmarks for the base model.

\paragrapht{Retrieval scale. }
To examine the impact of the number of retrieved audio chunks, we vary $k$, as shown in \cref{fig:figure3_a}.
All three models improve up to $k$=$15$, after which the trend becomes model-dependent, motivating our choice.

% Increasing $k$ consistently improves performance across all three models up to $k=15$, suggesting that retrieving a larger set of chunks provides additional evidence for multi-hop reasoning.
% Beyond $k=15$, however, the performance trends become model-dependent, with some models benefiting from additional chunks while others show limited gains or degradation.
% This suggests that retrieving more chunks does not necessarily provide additional useful evidence once a sufficient retrieval scope is reached.

\paragrapht{Unit length.}
\cref{fig:figure3_b} shows that the shortest 5s acoustic unit achieves the highest performance across models, showing that finer-grained acoustic units are sufficient for retrieval.

\paragrapht{Cost.}
In \cref{tab:inference_cost},
\ours achieves an $11.1\times$ and a $17.6\times$ reduction in latency and TFLOPs, respectively, compared to full audio.
% \revise{This demonstrates that \ours can substantially reduce the amount of audio processed at inference time without sacrificing performance, showing the benefit of retrieving only relevant evidence for long-form audio reasoning.}
Explicit planning in \cite{someki2026planrag} consumes $11.5\times$ times more than \ours for inference, while \cite{zhuang2026linearrag} uses less computation with text-only RAG but requires $1.6\times$ latency due to entity extraction for each query.
% \textcolor{blue}{Compared with PlanRAG-Audio, which is also designed for long-form audio understanding, \ours achieves substantially lower inference-time cost, reducing latency from 16.59s to 0.75s while also requiring fewer TFLOPs.}
Although \ours requires an indexing cost due to graph construction, this cost is incurred offline and only once per recording, taking just 391s for an average 32.9m recording.
% At inference time, \ours requires substantially less computation than processing the full audio while enabling multi-hop evidence retrieval over long-form audio.

\section{Conclusion}
We introduced \ours, a hierarchical \graph-based RAG framework for multi-hop reasoning over long audio recordings.
\ours simultaneously organizes textual and acoustic evidence into a hierarchical graph and propagates cross-modal relevance through compacted graph structure.
Experiments demonstrate consistent performance improvements with \ours across diverse LALM architectures, indicating that its benefits are not limited to any specific model.
Future work includes reducing the offline indexing cost through incremental graph updates, and extending retrieval to incorporate incoming information.

\bibliographystyle{IEEEbib}
\bibliography{ref}

\end{document}